\documentclass[12pt]{iopart}

\usepackage[dvipsnames]{xcolor}
\usepackage{graphicx}
\usepackage[
colorlinks,
citecolor=blue,
linkcolor=blue,
urlcolor=blue,plainpages=false,pdfpagelabels
]{hyperref}
\usepackage{txfonts}
\usepackage{iopams}
\usepackage{bbm}

\newcommand{\ket}[1]{\ensuremath{\,|{#1}\rangle}}
\newcommand{\braket}[1]{\ensuremath{\langle{#1}\rangle}}

\begin{document}
	\title[Generalized Chern numbers based on open system Green's functions]{Generalized Chern numbers based on open system Green's functions}

	\author{M. Bel\'en Farias, Solofo Groenendijk and Thomas L. Schmidt}
	\address{Department of Physics
	and Materials Science, University of Luxembourg, 1511 Luxembourg, Luxembourg}	
	
	\date{\today}
	
	\begin{abstract}
	
	We present an alternative approach to studying topology in open quantum systems, relying directly on Green's functions and avoiding the
	need to construct an effective non-Hermitian Hamiltonian. We define an energy-dependent Chern number based on the eigenstates
	of the inverse Green's function matrix of the system which contains, within the self-energy,
	all the information about the influence of the environment, interactions, gain or losses. We explicitly
	calculate this topological invariant for a system consisting of a single 2D Dirac cone and find that it is half-integer quantized
	when certain assumptions over the damping are made. Away from these conditions, which cannot or are not usually
	considered within the formalism of non-Hermitian Hamiltonians, we find that such a quantization is usually lost and the Chern number vanishes, and that in special cases, it can change to integer quantization.
	\end{abstract}


	\maketitle
	
	\section{Introduction} \label{sec:int}

To study the properties of quantum systems and understand how they manifest themselves in our macroscopic everyday world, it is usually necessary to take into account the interaction of such systems with their surroundings. In condensed-matter systems, the effects of an environment, be it the coupling to phonons, the presence of impurities~\cite{Mahan2000,bruus2004many}, or external processes that produce gain or loss \cite{Bruch2016,Haughian2018} are hard to avoid.
As an exact mathematical treatment of the degrees of freedom of both the system and its environment is often very complex, an open quantum system approach \cite{breuer2002book} is usually the best strategy to study and understand quantum systems in more realistic settings and exploit their unique properties.	

A relatively new direction is the study of topology in such open systems \cite{Diehl2011,Bardyn2013}. One approach that has gained traction lately
is the use of effective non-Hermitian (nH) Hamiltonians \cite{alvarez2018topological,Torres2019,song2019nonhermitian,bergholtz2019review,Rotter2009}.
In this formalism, the environment is accounted for by supplementing the Hamiltonian of the isolated system under study with non-Hermitian (nH) terms,
which can be thought of as arising from tracing out the degrees of freedom of the environment. In dissipative systems, the non-Hermitian part is related to the broadening of the system's energy levels \cite{Giusteri2015}, i.e., to the inverse life-time of the quasi-particle excitations  \cite{bergholtz2019review,alvarez2018topological}.

The nH framework has predicted intriguing phenomena such as the nH bulk-boundary correspondence \cite{Kunst2018}, the nH skin effect -- a feature of nH Hamiltonians in which the majority of the eigenstates are localized at the boundaries \cite{longhi2019probing,song2019nonhermitian} -- and the conversion of
band touching points into exceptional points at which two eigenstates coalesce and the Hamiltonian
becomes defective \cite{bergholtz2019review,moors2019disorder,carlstrom2018exceptional}.
A topological classification of the band structures of nH systems has been proposed in Refs.~\cite{Wojcik2020,shen2018topological}, and the Altland-Zirnbauer classification of topological invariants \cite{Altland1997} has been recently extended to nH systems \cite{kawabata2019symmetry,Gong2018,Liu2019}. However, in contrast to topological invariants in hermitian systems, those constructed from nH Hamiltonians are less obviously connected to physical quantities.

For example, in Ref.~\cite{shen2018topological}, Shen \textit{et al.}~have used the left and right eigenstates of a nH Hamiltonian to define a unique nH Chern number, which is quantized even though the Hall conductivity of the system is not. This result is at first surprising, because topological invariants of interacting systems were in fact defined already much earlier from the interacting Green's function and were used to prove the quantization of the Hall conductivity even in interacting quantum Hall systems \cite{Ishikawa1986,Ishikawa1987,wang2012simplified}. However, for generic nH Hamiltonians the Matsubara Green's functions used in these works become discontinuous, which makes this approach inapplicable and ultimately allows for a loss of quantization of the Hall conductivity in open systems despite the existence of a quantized nH Chern number \cite{hirsbrunner2019topology,philip2018loss,Groenendijk2020}.

The main goal of this work is to extend the range of open systems which can be classified topologically. For this purpose, we define a topological invariant directly from the Green's functions. Our approach is general in that we do not assume specific analytic properties of the Green's function. It thus applies to the retarded Green's function, which describes equilibrium properties in open systems, but also to the Keldysh Green's function, which can capture non-equilibrium dynamics. Our formalism does not apply to the Matsubara Green's function as it can become discontinuous.
A further advantage of using Green's functions is the possibility to account for more general environments. Concretely, the main difference between this approach and nH Chern numbers is that in the latter, the self-energy $\Sigma(\omega,\textbf{k})$ is usually assumed to be $\omega$-independent in order to obtain a time-local Hamiltonian formalism \cite{Giusteri2015,carlstrom2018exceptional,philip2018loss,kozii2017non,Yoshida2019}. This approximation is valid in many cases. In other cases, however, as for instance for quasi-particle spectra of disordered Weyl semimetals, the frequency dependence of the self-energy plays an important role \cite{moors2019disorder, zyuzin2018flat}. It is also worth noting that most studies involving nH Hamiltonians, particularly those interested in topological classifications, consider only momentum-independent nH terms. One notable exception to this rule is the recent preprint by Wang \textit{et al.}~which considers momentum-dependent decay rates \cite{wang21hall}.

A related topological invariant based on Green's function has been recently proposed by Kawabata \textit{et al.}~\cite{kawabata2020topological}. They showed that in one and three spatial dimensions, nH topological systems can be described by effective Chern-Simons field theories and are characterized by quantized topological response functions. The latter are in fact constructed from the Green's functions and depend on the energy at which the system is probed. However, this framework does not apply to two-dimensional open systems.

We shall present here a definition of such topological invariants in 2D systems, without having to rely on nH Hamiltonians. We will propose an analogous definition for a energy-dependent nH Chern number that can be obtained directly from the Green's functions, and show that this topological invariant is also quantized in many scenarios. Moreover, if this number is constructed based on the retarded Green's function and if certain assumptions about the energy and momentum dependence of the self-energy are fulfilled, we show that this invariant coincides with the nH Chern number defined in Ref.~\cite{shen2018topological}. Our definition, however, is more general and provides a topological invariant valid for a larger class of open systems. Moreover, we shall also delineate more precisely the conditions under which such quantization is lost, allowing us to shed some light onto the potential limitations of the nH topology.

\section{Non-Hermitian Chern numbers}

In particular, our work will extend the definition of a nH Chern number presented in Ref.~\cite{shen2018topological}.
There, the authors consider a nH Bloch Hamiltonian $\tilde{H}(k)$ with eigenstates
\begin{eqnarray}
\tilde{H} \ket{\psi^R_n} &= E_n \ket{\psi^R_n},\\
\tilde{H}^\dagger \ket{\psi^L_n} &= E^*_n \ket{\psi^L_n} \, .
\end{eqnarray}
where $n$ is the band index. Right and left eigenstates are distinct for nH Hamiltonians and are denoted by $\ket{\psi_n^{R,L}}$ here. These eigenstates then allow, a priori, the definition of Berry curvatures,
\begin{equation}
B_{n,ij}^{\alpha,\beta}(\textbf{k}) \equiv i \braket{\partial_i \psi_n^\alpha | \partial_j \psi_n^\beta} \, ,
\end{equation}
with $\alpha,\beta \in \{ L,R \}$ and $i,j \in \{ k_x, k_y \}$. The integral over the corresponding Berry curvature then leads to a topologically quantized nH Chern number.

An effective nH Hamiltonian can be thought of as consisting of two parts, $\tilde{H}=H+\Sigma$, where $H$
is the Hermitian Hamiltonian of the system without the influence of the environment or interactions, whereas $\Sigma$ is a nH correction to it.
Such a correction can often be interpreted as arising from the retarded Green's function,
\begin{equation}
G^{\rm ret}(\textbf{k},\omega)= \left[ \omega \mathbbm{1} - H(\textbf{k}) - \Sigma^{\rm ret}(\textbf{k}, \omega) \right]^{-1}   \, ,
\end{equation}
where $\mathbbm{1}$ is the identity matrix. For energy-independent $\Sigma^{\rm ret}$, we can see clearly that the eigenstates of the Green's function will coincide with the ones of the nH Hamiltonian (the eigenvalues will be different). However, for a general $\omega$-dependent self energy, this will no longer be the case. In this case, we can still calculate the eigenstates of the Green's function (or equivalently of its inverse) and thus define a new $\omega$-dependent Chern number which we will call $C(\omega)$. Since for $\omega = 0$, $C(\omega)$ coincides with the nH Chern number defined in Ref.~\cite{shen2018topological}, there will exist an interval close to $\omega = 0$ in which $C(\omega)$ will be quantized as well.

Hence, the first step in defining this new topological invariant is to study the eigenstates and eigenvalues
of the inverse Green's function. We shall do so in Sec.~\ref{sec:eig}. In Sec.~\ref{sec:chern} we will define the $\omega$-dependent Chern number and calculate
it for a specific model. We will show that, for certain types of damping, it is quantized. In Sec.~\ref{sec:generalenv} we generalize these results,
considering different situations for the influence of the environment, and find that such a quantization might change or be lost depending on the form of damping. Lastly, in Sec.~\ref{sec:conc} we present our conclusions.

\section{Eigenstates and eigenvalues of the inverse Green's function} \label{sec:eig}

\subsection{Eigenstates and biorthogonal basis}

As a first step, we are interested in calculating the eigenstates of the Green's function or equivalently of its inverse. To allow for a nontrivial topology, we assume a multiband system such that the Green's function becomes a matrix in the number of orbitals. So we are interested in diagonalizing the operator
\begin{equation} \label{eq:inv}
G^{-1}(\textbf{k},\omega)= \omega \mathbbm{1} - H(\textbf{k}) - \Sigma(\textbf{k},\omega) \, .
\end{equation}
If we had obtained the Green's function directly from the effective action of the system, it might be easier to calculate directly the eigenstates of $G(k,\omega)$ instead of those of the inverse. But in case of a Green's function constructed from the self-energy, it is more convenient to diagonalize the inverse~(\ref{eq:inv}).

As the simplest case of a nontrivial band topology, we will consider a two-band model, for which the most general form of the inverse Green's function can be cast as
\begin{equation}
G^{-1}(\textbf{k},\omega)= \epsilon(\textbf{k},\omega) \mathbbm{1} + \textbf{d}(\textbf{k},\omega) \cdot \boldsymbol{\sigma} \, ,
\end{equation}
where $\boldsymbol{\sigma}$ is the vector whose components are the three Pauli matrices. This operator is not Hermitian because
in general $\epsilon, d_i \in \mathbb{C}$. Its complex eigenvalues are given by
\begin{eqnarray} \label{eq:ev}
g_\pm (\textbf{k}, \omega) = \epsilon(\textbf{k}, \omega) \pm d(\textbf{k}, \omega)
\end{eqnarray}
where
\begin{eqnarray}
d= \sqrt{d_x^2+d_y^2+d_z^2} \, .
\end{eqnarray}
Note that $d$ is complex and in general does not coincide with the norm of $\textbf{d}$, the latter given by $\sqrt{|d_x|^2+|d_y|^2+|d_z|^2}$. The (unnormalized) eigenstates of $G^{-1}$, which fulfil $G^{-1} \ket{\chi_\pm^R} = g_\pm \ket{\chi_\pm^R}$, are of the form
\begin{eqnarray} \label{eq:chi}
\ket{\chi_\pm^R} =
\left(
\begin{array}{c}
d_z \pm d \\
d_x + i d_y
\end{array}
\right)
\, .
\end{eqnarray}
Since $G^{-1}$ is not a Hermitian operator, its right and the left eigenstates do not coincide. So we can define the left
eigenstates such that $(G^{-1})^\dagger \ket{\chi_\pm^L} = g^*_\pm \ket{\chi_\pm^L}$. A possible choice is
\begin{eqnarray}
\ket{\chi_\pm^L} =
\left(
\begin{array}{c}
d_z^* \pm d^* \\
d_x^* + i d_y^*
\end{array}
\right)
\, .
\end{eqnarray}
In general, the eigenstates $\lbrace \ket{\chi_\pm^R}, \ket{\chi_\pm^L} \rbrace$ form a biorthogonal basis \cite{brody2013biorthogonal}, as we
can readily check that
\begin{eqnarray}
\braket{\chi_\pm^L|\chi_\mp^R} & = \left( d_z \pm d, d_x - i d_y \right)
\left(
\begin{array}{c}
d_z \mp d \\
d_x + i d_y
\end{array}
\right)  = d_z^2 - d^2 + d_x^2 + d_y^2 = 0 \, .
\end{eqnarray}
However, for certain values of $d_{x,y,z}$, the eigenvectors $\ket{\chi^R_\pm}$ do not form a basis of the Hilbert space. Indeed, for $d_x = d_y = 0$ and $d_z = -d$ ($d_z = d$), the eigenvector $\ket{\chi^R_+}$ ($\ket{\chi^R_-}$) vanishes. Near this point, one therefore has to resort to a different choice of eigenvectors, for instance
\begin{eqnarray} \label{eq:chitilde}
\ket{\tilde{\chi}_\pm^R} = \left(
\begin{array}{c c}
d_x - i d_y \\
-d_z \pm d
\end{array}
\right) \,.
\end{eqnarray}
Hence, each of the four eigenstates $\ket{\chi_\pm^R}, \ket{\tilde{\chi}_\pm^R}$ can vanish at certain points in momentum space. This behavior is in fact essential for the definition of a Chern number \cite{shen2018topological}. Indeed, let us have a closer look at what happens when $d_x = d_y = 0$.

If $d_x = d_y = 0$,  we have $d = \sqrt{d_z^2}$. By looking at the definitions of the states, we can see that
if $d = d_z$, the states $\ket{\chi_-}$ and $\ket{\tilde{\chi}_+}$ vanish, while
if $d= - d_z$, $\ket{\chi_+}$ and $\ket{\tilde{\chi}_-}$ vanish. Let us now formulate in a more concrete way the conditions for which each of these eigenvectors vanish.
To do that, let us recall the definition of a complex square root \cite{Marsden1987}
\begin{eqnarray}
\sqrt{u + i v}= \pm \left(
\sqrt{\frac{u + \sqrt{u^2 + v^2}}{2}}
+ i \, \textrm{sgn}(v) \sqrt{\frac{-u + \sqrt{u^2 + v^2}}{2}}
\right) \, .
\end{eqnarray}
With this definition, we can easily see that
\begin{eqnarray} \label{eq:sqrt}
\sqrt{d_z^2}= d_z \textrm{sgn}\left[ \mathbb{R}\textrm{e}(d_z) \right] \, ,
\end{eqnarray}
which tells us that the sign of the real part of $d_z$ determines which is the well-defined set of eigenstates for the inverse Green's function.
We see that for $d_x = d_y = 0$, if $\mathbb{R}\textrm{e} (d_z)>0$,  $d = d_z$. While for $\mathbb{R}\textrm{e} (d_z) <0$ and $d_x = d_y = 0$, $d = - d_z$.

With this information we can finally write the complete set of eigenstates for our system.
These states shall be normalized following the convention for biorthogonal basis for which
we require that $\braket{\chi_\pm^L|\chi_\pm^R} = 1$ \cite{brody2013biorthogonal}. We then have

\begin{eqnarray} \label{eq:states}
\ket{\psi_+^R(\textbf{k},\omega)}_{\mathbb{R}e (d_z)< 0}=& \frac{1}{\sqrt{2 d(d - d_z)}}
\left(
\begin{array}{c}
d_x - i d_y \\
d - d_z
\end{array}
\right) \, ,   \\
\ket{\psi_-^R(\textbf{k},\omega)}_{\mathbb{R}e (d_z)< 0}=& \frac{1}{\sqrt{2 d(d - d_z)}}
\left( \begin{array}{c c}
d_z - d \\
d_x + i d_y
\end{array}
\right) \, , \\
\ket{\psi_+^R(\textbf{k},\omega)}_{\mathbb{R}e (d_z)> 0}=& \frac{1}{\sqrt{2 d(d + d_z)}}
\left(
\begin{array}{c}
d_z + d \\
d_x + i d_y
\end{array}
\right) \, ,   \\
\ket{\psi_-^R(\textbf{k},\omega)}_{\mathbb{R}e (d_z)> 0}=& \frac{1}{\sqrt{2 d(d + d_z)}}
\left( \begin{array}{c c}
d_x - i d_y \\
-d_z - d
\end{array}
\right)
\, .
\end{eqnarray}
The corresponding left eigenstates are obtained by changing $\textbf{d} \rightarrow \textbf{d}^*$. It is easy to see that, for a
given value of $d_z$, the set $\{ \ket{\psi^R_+},\ket{\psi^R_-},\ket{\psi^L_+}, \ket{\psi^L_-}\}$ forms a biorthogonal basis.

\subsection{Degeneracies, band crossings and exceptional points}

In order to be able to define a Chern number for our system, we need to guarantee that the system will remain \textit{separable} in the
sense that there are no touching points between the complex bands determined by the eigenvalues of the inverse Green's function.
These bands will cross in any point at which $g_+(\textbf{k},\omega)=g_-(\textbf{k},\omega)$, which could hold if and only if $d=0$. This extends the notion of separability of non-Hermitian Hamiltonians \cite{shen2018topological} to Green's functions.

It is worth noting that points in which the condition for the bands to cross is fulfilled, when $d=0$, are actually exceptional points
where the two eigenstates coalesce. For instance for $d=0$ one can see that $(d_x + i d_y) \ket{\psi_+^R(\textbf{k},\omega)}_{\mathbb{R}e (d_z)< 0} = -d_z \ket{\psi_-^R(\textbf{k},\omega)}_{\mathbb{R}e (d_z)< 0}$.
That is, the two states become linearly dependent, and a complete set of eigenstates can no longer be defined. The only exception to this
statement is the point $d_x = d_y = d_z = 0$. At this point the system becomes not only Hermitian, but also trivial (i.e., proportional
to the identity), and the states remain well-defined and linearly independent, so that this particular point is not an EP, but a Hermitian band crossing.

Let us briefly consider the specific system we shall use later on in this paper: a single 2D Dirac cone interacting with an environment that
has been integrated out, and whose effect on the system
will be encoded in four complex functions $\Gamma_\mu=(\Gamma_0,\Gamma_x,\Gamma_y,\Gamma_z)$, leading to a self-energy $\sum_{i=x,y,z}\Gamma_i \sigma_i + \Gamma_0 \sigma_0$. So we have
\begin{eqnarray} \label{eq:system}
\epsilon(\textbf{k},\omega) &= \omega - \Gamma_0 (\textbf{k},\omega)\, , \\
d_x(\textbf{k},\omega) &= - k_x v_x - \Gamma_x (\textbf{k},\omega)\, , \\
d_y(\textbf{k},\omega) &= - k_y v_y - \Gamma_y (\textbf{k},\omega)\, , \\
d_z(\textbf{k},\omega) &= -m - \Gamma_z (\textbf{k},\omega) \,  \label{eq:dz} .
\end{eqnarray}
The previous equations are in fact completely general and can describe any two-band model,
as long as the functions $\Gamma_\mu$ are arbitrary functions of $\omega$
and \textbf{k}.

The first conclusion we can draw is that the damping through the '0-channel', that is, the damping kernel that is proportional to the identity matrix,
can never close the gap, regardless of its functional form, and whether it develops an imaginary part or not.
Moreover, since we know that the nH Chern number is solely determined by the eigenstates, which do not depend on $\epsilon(\mathbf{k},\omega)$, any damping introduced through that channel won't affect the Chern number. As such, if it were the sole source of damping ($\Gamma_0\in \mathbb{C}$ but $\Gamma_i\in\mathbb{R}$), the Chern number would be indistinguishable from the Hermitian case ($\mathrm{Im}\Gamma_0=0$) and hence would be quantized to half integer \cite{Qi2011,Bernevig2013}.

Another particular case worth considering is where the imaginary part of only one of the three $\Gamma_i$ is non-vanishing. Consider for instance, $\Gamma_x, \Gamma_y \in \mathbb{R}$, but $\Gamma_z = i \gamma$ where $\gamma$ is a real constant. In such a case
\begin{eqnarray}
&d_x^2(\textbf{k},\omega) + d_y^2(\textbf{k},\omega) + d_z^2 (\textbf{k},\omega) = 0, \\
&(k_x v_x - \Gamma_x(\textbf{k},\omega))^2 + (k_y v_y - \Gamma_y(\textbf{k},\omega))^2 + m^2 - \gamma^2(\textbf{k},\omega)
+ 2 i m \gamma(\textbf{k},\omega) = 0,
\end{eqnarray}
and we can see from the last equation that, if the gap is originally open ($m \neq 0$), it will remain so even in the presence of damping. In this case the system will never encounter an exceptional point.

In contrast, in the case where $\Gamma_x = i\gamma$ but $\Gamma_{y,z} \in \mathbb{R}$, there are two exceptional points at the momenta $(k_x,k_y)= (0,  \pm\sqrt{\gamma^2-m^2}/v_y)$ if $|\gamma|>m$. In such a case, the Chern number would become ill-defined, so we will assume $|\Gamma_x| < m$ when considering the case of finite $\Gamma_x$. The case $\Gamma_y=i\gamma$ but $\Gamma_{x,z}\in \mathbb{R}$ is analogous.

\section{Berry curvature and Chern number} \label{sec:chern}

\subsection{Berry curvature and Chern number from Stoke's Theorem}

With the states defined in \Eref{eq:states} we can define the following $\omega$-dependent Berry curvature
\begin{eqnarray} \label{eq:Berry}
B_{\pm, ij} (\textbf{k},\omega) \equiv i \braket{\partial_i \psi_\pm^L (\textbf{k},\omega) | \partial_j \psi_\pm^R (\textbf{k},\omega)} \, ,
\end{eqnarray}
giving rise to an $\omega$-dependent Chern number defined as
\begin{eqnarray}
C_\pm(\omega)=\frac{1}{2\pi} \int d^2\textbf{k} \, \epsilon_{ij} \, B_{\pm, ij}^{LR} (\textbf{k},\omega) \, ,
\end{eqnarray}
where $\epsilon_{ij}$ denotes the Levi-Civita symbol in two dimensions and the summation over $i$ and $j$ is implied. The last expression can be rewritten as
\begin{eqnarray}
C_\pm(\omega)
=
\frac{i}{2 \pi}
\int d^2\textbf{k} &\left[ \epsilon_{i j}  \partial_i \braket{\psi_\pm^L (\textbf{k},\omega) | \partial_j \psi_\pm^R (\textbf{k},\omega)}
\right. \\
&  - \underbrace{\epsilon_{i j}  \braket{\psi_\pm^L (\textbf{k},\omega) |  \partial_i \partial_j \psi_\pm^R(\textbf{k},\omega)}}_{= 0 }
\Big] \, ,
\end{eqnarray}
where the last term vanishes because it is the contraction of a symmetric and an antisymmetric quantity.
In vectorial notation, we have
\begin{eqnarray}
C_\pm(\omega) &=
\frac{i}{2 \pi}
\int d\textbf{S} \cdot   \nabla \times \braket{\psi_\pm^L (\textbf{k},\omega) | \nabla \psi_\pm^R(\textbf{k},\omega)} \, ,
\end{eqnarray}
where $d\textbf{S} = d^2\textbf{k} \, \hat{z}$. The surface over which the integral is performed is, in principle, the whole $(k_x,k_y)$
plane ($\mathbb{R}^2$). We can consider it (going over to polar coordinates) as a disk of radius $\rho \rightarrow \infty$, and use Stokes's theorem to write the integral
as a line integral over the boundary $\mathcal{C}$ of this disk,
\begin{eqnarray}
C_\pm(\omega) &=
\frac{i}{2 \pi} \int_\mathcal{C} \underbrace{d\boldsymbol{\ell}}_{\rho d\theta \hat{\theta}} \cdot \braket{\psi_\pm^L (\rho,\theta,\omega) | \underbrace{\nabla}_{ \frac{1}{\rho} \partial_\theta \hat{\theta} + ... }| \psi_\pm^R(\rho,\theta,\omega)} \\
&= \lim_{\rho \rightarrow \infty}
\frac{i}{2 \pi} \int_0^{2 \pi} \, d\theta \,  \braket{\psi_\pm^L (\rho,\theta,\omega) | \partial_\theta \psi_\pm^R(\rho,\theta,\omega)} \, \label{eq:Chern}.
\end{eqnarray}
This last expression remains valid for any two-band model, and for any type of Green's function (advanced, retarded and Keldysh) except the Matsubara Green's function. As we mentioned earlier, even though a Chern number can be constructed from the Matsubara Green's function \cite{wang2012simplified}, it is, due to possible discontinuities, in general not a quantized topological invariant \cite{hirsbrunner2019topology}. In the following, we will specify a concrete system and explicitly calculate this non-Hermitian Chern number.

\subsection{Two-band model}

From now on, we will consider a single 2D Dirac cone interacting with an environment that has been already integrated out.
The effect of the environment is contained in the four $\Gamma_\mu$ functions, and the whole system
is defined by the expressions in Eqs.~(\ref{eq:system})-(\ref{eq:dz}).

In principle, these $\Gamma_\mu$ functions are arbitrary functions of $\omega$ and $\textbf{k}$. Depending on the symmetry of the system, some of them
may vanish while others remain nonzero, depending on the type of damping (or gain) present in the system. To be specific, we will begin by
considering the case when $\Gamma_z (\omega) \neq 0$ and $\Gamma_0 (\omega) \neq 0$, while $\Gamma_x = \Gamma_y = 0$.
This case corresponds for example to the self-energy arising from electron-phonon scattering to the lowest order in the coupling constants \cite{kozii2017non}.
Later on, we will extend our results to the case where $\Gamma_z$ is a function of momentum as well, and to the cases
where damping is present in other channels. We will also assume that the system has rotational symmetry, $v_x = v_y = v$, but we will relax this assumption later.

Under these assumptions, the parameters of the system become
\begin{eqnarray}
&\epsilon(\textbf{k},\omega) = \omega - \Gamma_0 (\omega)\, , \\
&d_x = - v \rho \cos \theta\, , \\
&d_y = - v \rho \sin \theta\, , \\
&d_z = - m - \Gamma_z(\omega)\, , \\
&d = \sqrt{v^2 \rho^2 + (m + \Gamma_z)^2} \, ,
\end{eqnarray}
where we have used polar coordinates such that $k_x = \rho \cos \theta$ and $k_y = \rho \sin \theta$. In this case, the choice of eigenstates presented in \Eref{eq:states} will be determined by $\textrm{sgn}[\mathbb{R}\textrm{e} (m+\Gamma_z(\omega)]$ .

\subsection{Quantized \texorpdfstring{$\omega$}{omega}-dependent Chern number}

We are now in a position to explicitly calculate the $\omega$-dependent Chern number defined in \Eref{eq:Chern}.
In the case considered here, with rotational symmetry and $\Gamma_z(\omega) \neq 0$, the eigenstates of the inverse
Green function take a simple form, namely:
\begin{eqnarray} \label{eq:psiRv}
\ket{\psi_+^R(\rho,\theta,\omega)}_{\mathbb{R}e (d_z)< 0}=& \frac{1}{\sqrt{2 d(d - d_z)}}
\left(
\begin{array}{c}
- v \rho e^{-i \theta}\\
d - d_z
\end{array}
\right) \, ,   \\
\ket{\psi_-^R(\rho,\theta,\omega)}_{\mathbb{R}e (d_z)< 0}=& \frac{1}{\sqrt{2 d(d - d_z)}}
\left( \begin{array}{c c}
d_z - d \\
- v \rho e^{i \theta}
\end{array}
\right) \, ,\\
\ket{\psi_+^R(\rho,\theta,\omega)}_{\mathbb{R}e (d_z)> 0}=& \frac{1}{\sqrt{2 d(d + d_z)}}
\left(
\begin{array}{c}
d_z + d \\
- v \rho e^{i \theta}
\end{array}
\right) \, ,   \\
\ket{\psi_-^R(\rho,\theta,\omega)}_{\mathbb{R}e (d_z)> 0}=& \frac{1}{\sqrt{2 d(d + d_z)}}
\left( \begin{array}{c c}
-v \rho e^{-i \theta} \\
-d_z - d
\end{array}
\right) \, , \label{eq:psiRvfinal}
\end{eqnarray}
where $d$ and $d_z$ are independent of $\theta$.

For each band ($\pm$), we have two different states, depending on the sign of $\mathbb{R}\textrm{e}(d_z)$. Since the shape of
the two states corresponding to each band is quite different, we will do the calculation independently.

For $C_+$ and $\mathbb{R}\textrm{e}(d_z)<0$, the derivative appearing in \Eref{eq:Chern} gives rise to
\begin{eqnarray}
\ket{\partial_\theta \psi_+^R(\rho,\theta,\omega)}_{\mathbb{R}e (d_z)< 0}=& \frac{1}{\sqrt{2 d(d - d_z)}}
\left(
\begin{array}{c}
i v \rho e^{-i \theta}\\
0
\end{array}
\right) \, ,
\end{eqnarray}
so that
\begin{eqnarray}
&C_+^{\mathbb{R}e (d_z)< 0}(\omega)
= \lim_{\rho \rightarrow \infty}
\frac{i}{2 \pi} \int_0^{2 \pi} \, d\theta \,  \frac{(- i v^2 \rho^2)}{2 d (d - d_z)} \\
&=  \lim_{\rho \rightarrow \infty} \frac{ v^2 \rho^2}{2 \sqrt{v^2 \rho^2 + (m + \Gamma_z)^2} (\sqrt{v^2 \rho^2 + (m + \Gamma_z)^2} + m + \Gamma_z)} = \frac{1}{2}
\, .
\label{eq:constGamma}
\end{eqnarray}
When $\mathbb{R}\textrm{e}(d_z)<0$, the derivative becomes
\begin{eqnarray}
\ket{\partial_\theta \psi_+^R(\rho,\theta,\omega)}_{\mathbb{R}e (d_z)> 0}=& \frac{1}{\sqrt{2 d(d + d_z)}}
\left(
\begin{array}{c}
0\\
- i v \rho e^{-i \theta}
\end{array}
\right) \, ,
\end{eqnarray}
and thus
\begin{eqnarray}
& C_+^{\mathbb{R}e (d_z)> 0}(\omega) = \lim_{\rho \rightarrow \infty}
\frac{i}{2 \pi} \int_0^{2 \pi} \, d\theta \,  \frac{i v^2 \rho^2}{2 d (d + d_z)} = - \frac{1}{2} \, .
\end{eqnarray}

Then, recalling that $d_z = - m - \Gamma_z(\omega)$, we summarize these two results as
\begin{eqnarray}
C_+(\omega) = \frac{1}{2} \textrm{sgn} \left[ \mathbb{R}\textrm{e}(m + \Gamma_z(\omega) ) \right] \, .
\end{eqnarray}

Repeating the same procedure for $C_-$, we arrive to the final expression for the $\omega$-dependent Chern number
in our model:
\begin{eqnarray} \label{eq:quantizedChern}
C_\pm(\omega) = \pm \frac{1}{2} \textrm{sgn} \left[ \mathbb{R}\textrm{e}(m + \Gamma_z(\omega) ) \right] \, .
\end{eqnarray}
We recall at this point that for a Hermitian single massive 2D Dirac cone, the Chern number is $C= \textrm{sgn}(m)/2$ \cite{Qi2011,Bernevig2013}. Our result for the $\omega$-dependent Chern number thus reflects the fact that the real part of $\Gamma_z$ can be interpreted as a renormalization of the mass $m$. So far, the assumption we made is that the influence of such an environment, i.e., the $\Gamma_\mu$ functions, depend only on $\omega$, and not on \textbf{k}. We have found that this topological invariant is half-quantized for all values of $\omega$, though it might flip the sign at some value depending on the explicit form of $\Gamma_z (\omega)$.
We have also found that the value of this $\omega-$dependent Chern number is independent of the imaginary part of $\Gamma_z$, that is, of
the damping.

In this Section we have explicitly calculated the $\omega$-dependent Chern number for the case with rotational symmetry $v_x = v_y = v$
since it provides us with the key results without getting into algebraically complicated steps.
However, these results can be extended to the case $v_x \neq v_y$ in a straightforward though cumbersome way,
and the Chern number calculated in this way is exactly the same shown in \Eref{eq:quantizedChern}.
In the
next Section, we will study how this result is affected for a more general damping: allowing for a dependence on the
momentum, and on different channels.

\section{\texorpdfstring{$\omega$}{omega}-dependent Chern number for more general environments} \label{sec:generalenv}

\subsection{The case in which \texorpdfstring{$\Gamma_z$}{Gammaz} depends on the wavevector}

So far we have considered the case in which $\Gamma_z(\omega)$ did not depend on \textbf{k}, and $\Gamma_{x}=\Gamma_y =0$,
motivated by the model for the damping proposed in Ref. \cite{kozii2017non}.
Now we will consider a more general case in which we will let $\Gamma_z=\Gamma_z(\omega, \textbf{k})$ depend on the momentum as well, in order to understand how its affects the Chern number.
This will give us further information into the limitations of the nH Hamiltonian formalism, where usually the $\Gamma_\mu$ are set to be constants.
For simplicity, we will again consider the case with rotational symmetry in which $v_x = v_y$, and write
\begin{eqnarray}
d_x &= - v \rho \cos \theta\, , \\
d_y &= - v \rho \sin \theta\, , \\
d_z &= - m - \Gamma_z(\omega, \rho, \theta)\, , \\
d &= \sqrt{v^2 \rho^2 + (m + \Gamma_z(\omega, \rho, \theta))^2} \, .
\end{eqnarray}
Then the states can again be written as in Equations (\ref{eq:psiRv})-(\ref{eq:psiRvfinal}),
but now $d$ and $d_z$ might depend on $\theta$ as well. Because of this, the derivative with respect to $\theta$ is of course much more complicated.
We will show only the calculations for the lower band and $\mathbb{R}\textrm{e}(d_z)<0$, since the other three cases are
completely analogous.
The derivative can be written as
\begin{eqnarray}
&\ket{\partial_\theta \psi_-^R(\rho,\theta,\omega)}_{\mathbb{R}e (d_z)< 0} = \nonumber \\
&- \left( 2 d ( d - d_z) \right)^{-\frac{3}{2}}
\left[ d - \frac{m+\Gamma_z}{d} (2 d - d_z) \right] \partial_\theta \Gamma_z
\left(
\begin{array}{c}
d_z - d \\
-v \rho e^{i \theta}
\end{array}
\right)  \\
&+ \left(2 d (d - d_z) \right)^{-\frac{1}{2}}
\left(
\begin{array}{c}
\partial_\theta \Gamma_z \left(\frac{m + \Gamma_z}{d} \right) \\
- i v \rho e^{ i \theta}
\end{array}
\right) \, , \nonumber
\end{eqnarray}
which leads to the following expression for the Chern number:
\begin{eqnarray}
&C_-^{\mathbb{R}e (d_z)< 0}(\omega) =
\lim_{\rho \rightarrow \infty} \frac{i}{2 \pi} \int_0^{2 \pi} \, d\theta \,
\Bigg\{
-  \frac{\partial_\theta \Gamma_z}{(2 d(d - d_z))^2}
 \left[ d - \frac{m+\Gamma_z}{d} (2 d - d_z) \right] \nonumber \\
 & \times
\left[
(d_z - d)^2
+
v^2 \rho^2
\right]
+    \frac{1}{2 d(d - d_z)}
\left[
\partial_\theta \Gamma_z (d_z - d)  \left(\frac{m + \Gamma_z}{d} \right)
+ i v^2 \rho ^2
\right]
\Bigg\}  \, .
\end{eqnarray}
One can proceed with an analytic calculation, by assuming that $\Gamma_z$ is a separable function of
$\rho$ and $\theta$ and can be written as $\Gamma_z (\rho, \theta) = f(\rho) g(\theta)$ for some arbitrary complex functions $f$ and $g$.
What we shall see is that the result will strongly depend on whether $f(\rho)$ grows rapidly with $\rho$, or not.
So let us analyze each case separately.

\subsubsection{Slowly growing \texorpdfstring{$f(\rho)$}{f(rho)}}

Let us first consider the case in which $f(\rho)$ does not grow with $\rho$, or grows very slowly, so that
\begin{eqnarray}\label{eq:rho_slow}
\lim_{\rho \rightarrow \infty} \frac{f(\rho)}{\rho} = 0\, .
\end{eqnarray}
In this case, $d_z / \rho$, $f(\rho)/d$, and $d_z/d$ all vanish when $\rho \rightarrow \infty$, while $d / \rho \rightarrow |v|$.
So it's easy to see that
\begin{eqnarray}
C_-^{\mathbb{R}e (d_z)< 0}(\omega) &= \lim_{\rho \rightarrow \infty} \frac{i}{2 \pi} \int_0^{2 \pi} \, d\theta \,
\left[
-  \frac{f(\rho)}{4d} g'(\theta) \,  \right.
 \left\lbrace 1 - \left(\frac{m}{d} + \frac{f(\rho)}{d} g(\theta)\right) \left(2  - \frac{d_z}{d}\right) \right\rbrace
 \nonumber \\
&\times \left\lbrace
1
+
\frac{v^2 \rho^2}{(d - d_z)^2}
\right\rbrace
+ \left.
\frac{f(\rho)}{2d} g'(\theta)  \left(\frac{m}{d} + \frac{f(\rho)}{d} g(\theta) \right)
+  \frac{i v^2 \rho ^2 }{2 d(d - d_z)}
\right] \nonumber \\
&=  \frac{i}{2 \pi} \int_0^{2 \pi} \, d\theta \,
  \frac{i v^2 }{2 v^2} = - \frac{1}{2} \, .
\end{eqnarray}
Which is the same result obtained for $\Gamma_z$ independent of \textbf{k}, so we retain the half-quantization we have encountered. It is
easy to see that this result is independent of whether $\Gamma_z$ depends on $\theta$ or not. The condition (\ref{eq:rho_slow}) thus determines the range of validity of the approximation of treating the self-energy as $\textbf{k}$-independent in nH Hamiltonian.

\subsubsection{Rapidly growing \texorpdfstring{$f(\rho)$}{f(rho)}}

Let us now consider the case of a rapidly growing $f(\rho)$, such that
\begin{eqnarray}
\lim_{\rho \rightarrow \infty} \frac{f(\rho)}{\rho} = \infty \, .
\end{eqnarray}
In this case, we can see that $d_z/\rho \rightarrow \infty$ and $d/\rho \rightarrow \infty$ when $\rho \rightarrow \infty$, as well as
\begin{eqnarray}
& \lim_{\rho \rightarrow \infty} \frac{f(\rho)}{d} = \frac{1}{ \sqrt{g^2( \theta)}}  = \frac{1}{g \, \textrm{sgn} (\mathbb{R}\textrm{e}(g))}\, ,\\
& \lim_{\rho \rightarrow \infty} \frac{d_z}{d} = - \frac{g}{\sqrt{g^2}} = \frac{- g}{g \, \textrm{sgn} (\mathbb{R}\textrm{e}(g))} =- \textrm{sgn} (\mathbb{R}\textrm{e}(g))\, .
\end{eqnarray}
To calculate these limits, we have assumed that $f \in \mathbb{R}$ but $g \in \mathbb{C}$, and made use of \Eref{eq:sqrt}. On the other hand, as $f(\rho) \rightarrow \infty$ when $\rho \rightarrow \infty$, we can see that both $d, d_z \rightarrow \infty$.\\
All these considerations allow us to take the limit $\rho \rightarrow \infty$ in the Berry curvature
\begin{eqnarray}
\lim_{\rho \rightarrow 0} \braket{\psi_-^L(\rho,\theta,\omega) |\partial_\theta \psi_-^R(\rho,\theta,\omega)}_{\mathbb{R}e (d_z)< 0} = \frac{g'(\theta)}{g(\theta)} \, .
\end{eqnarray}

With this result we can write the Chern number as
\begin{eqnarray}\label{eq:Gammaz_rapid}
C_-^{\mathbb{R}e (d_z)< 0}(\omega)
& =  \frac{i}{2 \pi} \int_0^{2 \pi} \, d\theta \,
\frac{g'(\theta)}{g(\theta)}
=
\frac{i}{2 \pi}  \left\lbrace\log \left[ g(2 \pi) \right] - \log \left[g(0)\right] \right\rbrace \, .
%
\end{eqnarray}
From the last expression, we can see that this quantity is always quantized. In the case of real $g$ we see that the Chern number vanishes. However, in some other cases, the Chern number can be nontrivial, e.g., in the case $\Gamma_z = f(\rho) g(\theta) = f(\rho) e^{i \theta}$ one finds $C_-^{\mathbb{R}e (d_z)< 0}(\omega)= -1$. Hence, in this case, the presence of a self-energy can give rise to an \emph{integer} Chern number in a system where the Chern number only takes the values $\pm 1/2$ in the absence of the self-energy. For the retarded self-energy, such a $\Gamma_z(\rho,\theta)$ is not possible because the retarded self-energy must have a negative imaginary part for all $\theta$. For the Keldysh self-energy, in contrast, such a constraint does not apply, so such a nontrivial change of Chern number from half-integral quantized to integer quantized may arise in nonequilibrium systems.

With this result we come to
the conclusion that the half-quantization found in the cases where $\Gamma_z$ does not depend on $\rho$ or grows slowly with it,
does not hold in the case of rapidly growing $\Gamma_z$.
In this case, the Chern number is found to be quantized to an integer.
The case of rapidly growing $\Gamma_z$ is likely to be the quite common when considering realistic models for the environment, since the $\Gamma_\mu$ functions can be thought of as the decay rates of quasiparticles. Since the latter should be small compared to the quasiparticle energy for small \textbf{k}, the self-energies usually grow with a higher power of $\textbf{k}$ than linearly. As we showed above, in most cases the Chern number will then vanish, revealing that in these cases the damping renders the topology of the system trivial.

\subsection{Results for damping in a different channel}

Lastly we will extend our results to the case in which the damping, instead of acting in the $z$ channel, is in the $x$ channel. Due to symmetry, this is equivalent to taking it in the $y$ channel. Then, our system will be defined by
\begin{eqnarray}
&\epsilon(\textbf{k},\omega) = \omega - \Gamma_0 (\omega)\, , \label{eq:energy} \\
&d_x = - v \rho \cos \theta - \Gamma_x\, ,\\
&d_y = - v \rho \sin \theta\, , \\
&d_z = - m  \, ,
\end{eqnarray}
where
\begin{eqnarray}
d &= \sqrt{v^2 \rho^2 + m^2 + \Gamma_x^2 + 2 v \Gamma_x \rho \cos \theta} \,
\end{eqnarray}
and $\Gamma_x = \Gamma_x(\omega, \rho, \theta)$. Note that in this case, we have to make the additional assumption that $|\Gamma_x| < m$ for all $\textbf{k}$ and $\omega$ to avoid exceptional points.

We will again calculate only one of the four possible Chern numbers, since the remaining three are obtained
in an analogous way.
This time, let us have a look at $C_+^{\mathbb{R}e (d_z)< 0}(\omega)$ (which, in the present case, corresponds to
$m>0$).
The right $+$ eigenstate of the Green's function is then written as
\begin{eqnarray}
\ket{\psi_+^R(\rho,\theta,\omega)}_{\mathbb{R}e (d_z)< 0}=& \frac{1}{\sqrt{2 d(d - m)}}
\left(
\begin{array}{c}
- v \rho e^{-i \theta} - \Gamma_x \\
d - m
\end{array}
\right) \, ,
\end{eqnarray}
and its derivative with respect to $\theta$ is given by
\begin{eqnarray}
&\ket{\partial_\theta \psi_+^R(\rho,\theta,\omega)}_{\mathbb{R}e (d_z)< 0}
=
\frac{- (4 d - 2 m) \partial_\theta d}{2 (2 d^2 - 2 m d)^{3/2}}
\left(
\begin{array}{c}
- v \rho e^{-i \theta} - \Gamma_x \\
d - m
\end{array}
\right) \nonumber \\
&+
\frac{1}{\sqrt{2 d( d - 2 m)}}
\left(
\begin{array}{c}
i v \rho e^{-i \theta} - \partial_\theta \Gamma_x \\
\partial_\theta d
\end{array}
\right) \, ,
\end{eqnarray}
where
\begin{eqnarray}
\partial_\theta d =
\frac{2 \Gamma_x \partial_\theta \Gamma_x + 2 v \rho (\partial_\theta \Gamma_x \cos \theta - \Gamma_x \sin \theta )}
{2 \sqrt{v^2 \rho^2 + \Gamma_x^2 + m^2 + 2 v \Gamma_x \rho \cos \theta}} \, ,
\end{eqnarray}
and both $\Gamma_x$ and $\partial_\theta \Gamma_x$ depend on the explicit model for the damping.
The Berry curvature then reads
\begin{eqnarray}
&\braket{\psi_+^L(\rho,\theta,\omega) |\partial_\theta \psi_+^R(\rho,\theta,\omega)}_{\mathbb{R}e (d_z)< 0} = \\
&=  - \frac{(2 d -  m) \partial_\theta d}{ [ 2 d(d-m)]^2}
\left\lbrace v^2 \rho^2 + \Gamma_x^2 + 2 v \rho \cos \theta + (d - m)^2 \right\rbrace  \nonumber \\
&+ \frac{\partial_\theta d (d-m) - i v^2 \rho^2 + \Gamma_x \partial_\theta \Gamma_x
	+ \partial_\theta \Gamma_x v \rho e^{i \theta} - i \Gamma_x v \rho e^{- i \theta}}{2 d (d - m)}
\nonumber \, .
\end{eqnarray}

As we did when the damping was allocated in the $z$ channel, we will first consider the case in which $\Gamma_x(\omega)$
does not depend on \textbf{k}, and then see how such a dependence affects the results.

\subsubsection{\texorpdfstring{$\Gamma_x$}{Gammaz} independent of \texorpdfstring{\textbf{k}}{k}}

In this case, the limit $\rho \rightarrow \infty$, necessary to calculate the Chern number, becomes straightforward.
To take such limit it is useful to see that
\begin{eqnarray}
\lim_{\rho \rightarrow \infty} \frac{d}{\rho} &= \lim_{\rho \rightarrow \infty} \frac{\sqrt{v^2 \rho^2 + m^2 + \Gamma_x^2 + 2 v \Gamma_x \rho \cos \theta}}{\rho} = |v|\,  \\
\lim_{\rho \rightarrow \infty} \frac{\partial_\theta d}{\rho} &= \lim_{\rho \rightarrow \infty}  \frac{ - 2 v \rho \Gamma_x \sin \theta }
{2 \rho \sqrt{v^2 \rho^2 + \Gamma_x^2 + m^2 + 2 v \Gamma_x \rho \cos \theta}}  = 0 \, .
\end{eqnarray}
The Chern number, then, reads
\begin{eqnarray}
&C_+^{\mathbb{R}e (d_z)< 0}(\omega)
=  \lim_{\rho \rightarrow \infty} \frac{i}{2 \pi} \int_0^{2 \pi} d\theta
\left[ - \frac{(4 d - m) \partial_\theta d}{2 [ 2 d(d-m)]^2} \right. \nonumber \times \\
&\times \left\lbrace v^2 \rho^2 + \Gamma_x^2 + 2 v \rho \cos \theta + (d - m)^2 \right\rbrace \\
&+ \left. \frac{1}{2 d (d - m)}
\left\lbrace \partial_\theta d (d-m) - i v^2 \rho^2 - i \Gamma_x v \rho e^{- i \theta} \right\rbrace \right] = \frac{1}{2} \, \nonumber .
\end{eqnarray}
This result shows that the Chern number remains half-quantized when the damping is allocated in a different channel ($\Gamma_x \neq 0$, $\Gamma_z = \Gamma_y = 0$, or $\Gamma_y \neq 0$ and $\Gamma_z = \Gamma_x = 0$).

\subsubsection{\texorpdfstring{$\Gamma_x$}{Gammax} growing slowly with \texorpdfstring{$\rho$}{rho}}

If now we allow $\Gamma_x$ to be a function of both frequency and momentum, we have to again consider two different cases,
depending on its behaviour with $\rho$.
First we consider a slowly growing (or decreasing) $\Gamma_x = f(\rho) g(\theta)$, such that $f(\rho)/\rho \rightarrow 0$ when $\rho \rightarrow \infty$.\\
Then the Berry curvature reads:
\begin{eqnarray}
&\braket{\psi_+^L(\rho,\theta,\omega) |\partial_\theta \psi_+^R(\rho,\theta,\omega)}_{\mathbb{R}e (d_z)< 0} = \\
&=  - \frac{(2 d -  m) \partial_\theta d}{ [ 2 d(d-m)]^2}
\left\lbrace v^2 \rho^2 + f^2 g^2 + 2 v \rho \cos \theta + (d - m)^2 \right\rbrace \nonumber \\
&+ \frac{\partial_\theta d (d-m) - i v^2 \rho^2 + f^2 g g'  + f g' v \rho e^{i \theta}  -  i f g  v \rho e^{- i \theta}}{2 d (d - m)}\, ,
\nonumber
\end{eqnarray}
where
\begin{eqnarray}
d &= \sqrt{v^2 \rho^2 + m^2 + f^2 g^2 + 2 v \Gamma_x \rho \cos \theta}\, , \\
\partial_\theta d &=
\frac{2 f^2 g g' + 2 v \rho f (g' \cos \theta - g \sin \theta )}
{2 \sqrt{v^2 \rho^2 + f^2 g^2 + m^2 + 2 v f g \rho \cos \theta}} \, .
\end{eqnarray}
To take the limit $\rho \rightarrow \infty$, we use
\begin{eqnarray}
\lim_{\rho \rightarrow \infty} \frac{d}{\rho} &= |v|\,  \\
\lim_{\rho \rightarrow \infty}\frac{ \partial_\theta d}{\rho} &= 0 \, .
\end{eqnarray}
With these results, it is easy to see that
\begin{eqnarray}
C_+^{\mathbb{R}e (d_z)< 0}(\omega)
=&  \lim_{\rho \rightarrow \infty} \frac{i}{2 \pi} \int_0^{2 \pi} d\theta
\braket{\psi_+^L(\rho,\theta,\omega) |\partial_\theta \psi_+^R(\rho,\theta,\omega)}
= \frac{1}{2} \, ,
\end{eqnarray}
and we recover the half-quantized Chern number, for any $\theta$ dependence, as long as $\Gamma_x$ grows slowly with $\rho$.

\subsubsection{Rapidly growing \texorpdfstring{$\Gamma_x$}{Gammax}}

Now we take again $\Gamma_x = f(\rho) g(\theta)$, but this time we have $f(\rho)/\rho \rightarrow \infty$ when $\rho \rightarrow \infty$.
In this case, we get the following limits:
\begin{eqnarray}
\lim_{\rho \rightarrow \infty} \frac{d}{\rho} &= \frac{1}{g \, \textrm{sgn} (\mathbb{R}e (g) )}\, , \\
\lim_{\rho \rightarrow \infty}\frac{ \partial_\theta d}{\rho} &= \frac{g'}{g}\, ,  \\
\lim_{\rho \rightarrow \infty} \frac{\rho}{d} &= 0 \, .
\end{eqnarray}
Then, we can see that the Berry curvature in the limit of large $\rho$ vanishes:
\begin{eqnarray}
&\lim_{\rho \rightarrow \infty} \braket{\psi_+^L(\rho,\theta,\omega) |\partial_\theta \psi_+^R(\rho,\theta,\omega)}_{\mathbb{R}e (d_z)< 0} = \\
&=  \lim_{\rho \rightarrow \infty} \left[ - \frac{(2 -  \frac{m}{d}) \frac{\partial_\theta d}{d}}{ [ 2 (1-\frac{m}{d})]^2}
\left\lbrace v^2 \frac{\rho^2}{d^2} + \frac{f^2}{d^2} g^2 + 2 \frac{v}{d} \frac{\rho}{d} \cos \theta + (1 - \frac{m}{d})^2 \right\rbrace \right. \nonumber \\
&+ \left.  \frac{1}{2  (1 - \frac{m}{d})}
\left\lbrace \frac{\partial_\theta d}{d} (1-\frac{m}{d}) - i v^2 \frac{\rho^2}{d^2} + \frac{f^2}{d^2} g g' + \frac{f}{d} g' \frac{v}{d} \rho e^{i \theta} - i \frac{f}{d} g  v \frac{\rho}{d}  e^{- i \theta} \right\rbrace \right] \nonumber \\
&=  - \frac{1}{2} \frac{g'}{g}
\left( \frac{1}{g^2 \, \textrm{sgn}^2 (\mathbb{R}\textrm{e} (g) )}  g^2 + 1 \right)+
\frac{1}{2 } \left\lbrace
\frac{g'}{g}  + \frac{g g'}{g^2 \textrm{sgn}^2 (\mathbb{R}\textrm{e} (g) )} \right\rbrace= 0 \nonumber  \, ,
\end{eqnarray}
which implies that the Chern number vanishes as well. The main difference between this result and Eq.~(\ref{eq:Gammaz_rapid}) is that,
when the damping was allocated in the $\Gamma_z$ channel, certain functional forms of $\Gamma_z(\rho,\theta)$ can result in a non-vanishing, integer-quantized Chern number. In the case in which the damping is allocated in the $\Gamma_x$ or $\Gamma_y$
channels, however, this does not occur and the Chern number vanishes for any functional form of $g(\theta)$.

In any case, several conclusions can be extracted from the cases analyzed through this section. We can see that in many cases (when $\Gamma_i$
does not depend on \textbf{k}, or when it doesn't grow rapidly with $\rho$), we recover the half-quantized result that was observed
while using nH Hamiltonian formalism. In this sense, we see that our approach is robust, and consistent with the use of nH Hamiltonians. But on the other hand, we also
found that for some specific behaviour of the $\Gamma_i$ functions, this half-quantization is lost, which shows that the nH Hamiltonian formalism
might have some limitations for specific environments.

\section{Conclusions} \label{sec:conc}

In this work we have presented an alternative approach for the study and topological classification of open quantum systems. Throughout this paper we have investigated a system interacting with an environment whose influence on the system is encoded in a general self-energy. Moreover, such self-energies can arise not only from the coupling to an environment but also as an effective description of effects such as interactions which cannot be treated exactly. Such systems have often been topologically classified using effective nH Hamiltonians and we have proposed Green's function as an alternative classification method.

By relying on the system's Green's function instead of the construction of a non-Hermitian (nH) Hamiltonian, we render our approach more general and applicable in principle to any gapped 2D model. The formalism applies to advanced, retarded, as well as Keldysh Green's functions, and can thus capture effects due to environments, non-equilibrium, interactions, gain and losses, which can be expressed through the self-energy.

Using the eigenstates of the inverse Green's function, we were able to define a topological invariant, the $\omega$-dependent Chern number. This quantity is analogous to the nH Chern number defined in the literature, with two main differences: firstly, it does not require the construction of a nH Hamiltonian and hence the assumptions implied by such a construction. Secondly, this topological invariant is energy-dependent, and the value of this Chern number can indeed change at certain energies, even in the case in which it is quantized or half-quantized. This is in line with recent field-theoretical approaches which have also argued that since nH phases arise in open systems or systems out of equilibrium, the momentum and frequency degrees of freedom have to be treated on different footings \cite{kawabata2020topological}.

We have considered then in our work a concrete system: a continuum 2D Dirac model, consisting of a single gapped Dirac cone.
The influence of the environment, damping,
interactions, gain, losses, etc., was encoded within four functions (damping channels) denoted $\Gamma_\mu=(\Gamma_0,\Gamma_x,\Gamma_y,\Gamma_z)$. One of the results of our work is that the outcome significantly depends on the damping channel. Our first
result is that the damping allocated in the $\Gamma_0$ channel, which can be thought of as a (complex) renormalization of the energy
(see \Eref{eq:energy}), does not impact the Chern number, independently of its functional form.
We have considered the case in which the $\Gamma_\mu$ functions depend solely on $\omega$, being independent of \textbf{k}. If
only one of the $\Gamma_i$ with $i=x,y,z$ was non-vanishing, the Chern number is half-quantized to values $\pm 1/2$. The sign
is determined by the band and the sign of the mass. Since the mass is effectively renormalized by $\Gamma_z$, we found
that its real part might induce a jump in the Chern number. Since the $\Gamma_i$ functions are frequency dependent, so is the change in the Chern number. The Chern number might be different at different energies.

When we allowed the $\Gamma_\mu$ functions to depend as well on the momentum, we found that the results greatly depended on
the behaviour of $\Gamma_\mu$ at large momentum. For slowly-growing functions, the half-quantized result was recovered.
In the case of rapidly growing $\Gamma_\mu$, however, we found that in most cases the $\omega$-dependent Chern number vanished. Since the
$\Gamma_\mu$ functions can be interpreted as the decay rates of quasiparticles, this result
can be understood as the damping-induced loss of topological quantization.
We found, however, that for some particular functional forms of $\Gamma_z$, a non-vanishing, integer-quantized Chern number
could be obtained.

In summary, our results go beyond what has been considered so far in terms of topology of open quantum systems, particularly for
gapped two-band systems. We have made fewer assumptions on the influence of the environment on the system, and we have found that certain environment properties that cannot easily be taken into account when using the nH Hamiltonian formalism, might indeed produce significant changes to the topology of the system.
We believe our results will help to broaden the understanding we have so far of topology in open systems, shedding some light into
the regions of validity and the implied assumptions that underlie the use of nH Hamiltonians, as well as allowing for a broader range of systems
to be studied.

\section*{Acknowledgements}

We acknowledge financial support from the National Research Fund Luxembourg under grants ATTRACT A14/MS/7556175/MoMeSys, CORE C16/MS/11352881/PARTI, and CORE C20/MS/14757511/OpenTop

\section*{References}
\bibliographystyle{iopart-num}
\bibliography{topoenvironment}

\end{document}